\documentclass[a4paper]{spie}  

\usepackage[]{graphicx}
\usepackage{amssymb,amsmath,psfrag,url}

\newcommand{\curl}{\mathrm{curl}}
\newcommand{\Field}[1]{{\boldsymbol{#1}}}

\newcommand{\Kvec}[1]{{\vec{#1}}}
\newcommand{\dd}[1]{{\rm{d}}#1}
\newcommand{\SSS}[1]{#1}
\newcommand{\VSS}[1]{#1}
\newcommand{\bhp}{\SSS{H}^{1}_{\Kvec{k}}}
\newcommand{\bhc}{\VSS{H}_{{\Kvec{k}}}\left(\curl\right)}
\newcommand{\bhph}{\SSS{V}_{h, \Kvec{k}}}
\newcommand{\bhch}{\VSS{W}_{h, {\Kvec{k}}}}

\newcommand{\bhcP}{\VSS{H}^{\perp}_{{\Kvec{k}}}\left(\curl\right)}

\newcommand{\cnum}{\mathbb{C}}
\newcommand{\MyMatrix}[1]{\mathrm{#1}}
\newcommand{\MyVector}[1]{\mathrm{#1}}

\newtheorem{problem}{Problem}

\title{FEM Modelling of 3D Photonic Crystals and Photonic Crystal Waveguides} 

\author{
Sven Burger\supit{\,ab}, 
Roland Klose\supit{\,a}, 
Achim Sch{\"a}dle\supit{\,a},
Frank Schmidt\supit{\,ab}, and 
Lin Zschiedrich\supit{\,ab}
\skiplinehalf
\supit{a}
Zuse Institute Berlin,
Takustra{\ss}e 7,
D\,--\,14\,195 Berlin,
Germany
\skiplinehalf
\supit{b}
JCMwave GmbH,
Haarer Stra{\ss}e 14a,
D\,--\,85\,640 Putzbrunn, 
Germany
}

\authorinfo{
Further author information: (Send correspondence to S.B.)\\
URL: http://www.zib.de/nano-optics/\\
S.B.: E-mail: burger@zib.de, Telephone: +49 30 84185 302
}

\begin{document} 
  \maketitle 

\begin{abstract}
We present a finite-element simulation tool for calculating 
light fields in 3D nano-optical devices.
This allows to solve challenging problems on a standard personal 
computer.
We present solutions to eigenvalue problems, like Bloch-type
eigenvalues in photonic crystals and photonic crystal waveguides,
and to scattering problems, like the transmission through finite
photonic crystals.

The discretization is based on unstructured tetrahedral grids with 
an adaptive grid refinement controlled and steered by an error-estimator.
As ansatz functions we use higher order, vectorial elements 
(Nedelec, edge elements). 
For a fast convergence of the solution we make use of advanced multi-grid
algorithms adapted for the vectorial Maxwell's equations. 
\end{abstract}

\keywords{3D photonic crystals, photonic crystal waveguides, 
finite-element method, Maxwell's equations, nano-optics}

\section{Introduction}
Photonic crystals (PhC's)
are structures composed of different optical transparent materials 
with a spatially periodic
arrangement of the refractive index~\cite{Joannopoulos1995a,Sakoda2001a}.
Propagating light with a wavelength of the order 
of the periodicity length of the photonic crystal is significantly influenced
by multiple interference effects.
The most prominent effect is the opening of photonic bandgaps, in analogy to 
electronic bandgaps in semiconductor physics or atomic bandgaps in 
atom optics. 
Due to the fast progress in nano-fabrication technologies PhC's can be
manufactured with high accuracy.
This allows for the miniaturization of optical components 
and a broad range of technological applications, like, e.g., 
in telecommunications~\cite{Maerz2004a}. 
The properties of light propagating in PhC's are in general critically 
dependent on system parameters.
Therefore, the design of photonic crystal devices calls for 
simulation tools with high accuracy, speed and reliability.

\section{Light Propagation in Photonic Crystals}
Light propagation in photonic crystals is governed by Maxwell's equations
with the assumption of vanishing densities of free charges and currents.
The dielectric coefficient $\varepsilon(\vec{x})$ and the permeability 
$\mu(\vec{x})$ are real, positive and periodic, 
$\varepsilon \left(\vec{x}\right)  =  \varepsilon \left(\vec{x}+\vec{a} \right)$, 
$\mu \left(\vec{x} \right)  =  \mu \left(\vec{x}+\vec{a} \right)$.
Here $\vec{a}$ is any elementary vector of the crystal 
lattice~\cite{Sakoda2001a}.  For given primitive lattice vectors 
$\vec{a}_{1}$, $\vec{a}_{2}$ and $\vec{a}_{3}$ the elementary cell 
$\Omega\subset\mathbb R^{3}$ is defined as
$\Omega = \left\{\vec{x} \in \mathbb R^{3}\,|\,
x=\alpha_{1}\vec{a}_1+\alpha_{2}\vec{a}_2+\alpha_{3}\vec{a}_3;
0\leq\alpha_{1},\alpha_{2},\alpha_{3}<1
\right\}.$
A time-harmonic ansatz with frequency $\omega$ and magnetic field 
$\Field{H}(\vec{x},t)=e^{-i\omega t}\Field{H}(\vec{x})$ leads to
an eigenvalue equation for $\Field{H}(\vec{x})$ with the constraint that
$\Field{H}(\vec{x})$ is divergence-free:
\begin{equation}
\label{eigeneqn1}
\nabla\times\frac{1}{\varepsilon(\vec{x})}\,\nabla\times\Field{H}(\vec{x})
= \omega^2 \mu(\vec{x})\Field{H}(\vec{x}), \quad
\nabla\cdot\mu(\vec{x})\Field{H}(\vec{x}) = 0,
\qquad\vec{x}\in\Omega.
\end{equation}
Similar equations are found for the electric field 
$\Field{E}(\vec{x},t)=e^{-i\omega t}\Field{E}(\vec{x})$:
\begin{equation}
\label{eigeneqn1e}
\nabla\times\frac{1}{\mu(\vec{x})}\,\nabla\times\Field{E}(\vec{x})
= \omega^2 \varepsilon(\vec{x})\Field{E}(\vec{x}), \quad
\nabla\cdot\varepsilon(\vec{x})\Field{E}(\vec{x}) = 0,
\qquad\vec{x}\in\Omega.
\end{equation}

The Bloch theorem applies for wave propagation in periodic media.
Therefore we aim to find Bloch-type eigenmodes~\cite{Sakoda2001a} to 
Equations~(\ref{eigeneqn1}), 
defined as
\begin{equation}\label{Bloch1}
\Field{H}(\vec{x}) = e^{i \Kvec{k}\cdot\vec{x}} \Field{u}(\vec{x}), \qquad
\Field{u}(\vec{x})=\Field{u}(\vec{x}+\vec{a}).
\end{equation}
where the Bloch wavevector $\Kvec{k}\in\mathbb{R}^3$ is chosen from the first 
Brillouin zone.
A similar procedure yields the Bloch-type eigenmodes to Equations~(\ref{eigeneqn1e}),
however, in what follows we will concentrate on Equations~(\ref{eigeneqn1}).

In order to reformulate Equations~(\ref{eigeneqn1}) and~(\ref{Bloch1}) we define the following 
functional spaces and sesquilinear forms:
{\it \newline 
{\bf(a)} The set of Bloch periodic smooth functions is defined as
\[
C^{\infty}_{\Kvec{k}}\left(\Omega,\cnum^{d}\right) = 
\left\{w \in C^{\infty}\left(\Omega,\cnum^{d}\right)\;|\;
w\left(\vec{x}+\vec{a}\right) = 
e^{i\Kvec{k}\cdot\vec{a}}w\left(\vec{x}\right)\right\}.
\]
The Sobolev space $\bhc$ is the closure of 
$C^{\infty}_{\Kvec{k}}\left(\Omega,\cnum^{3}\right)$ with respect to the 
$\VSS{H}\left(\curl\right)$-norm. The space $\bhp$ 
is defined accordingly. 
\newline
{\bf (b)} The sesquilinear forms $a : \bhc\times\bhc \rightarrow\mathbb{C}$ and
$b :  \bhc\times\bhc \rightarrow\mathbb{C}$ are defined as
\begin{align}
a\left(\Field{w},\Field{v}\right)&=\int_\Omega \frac{1}{\varepsilon}(\nabla 
\times \Field{w}) 
\cdot \overline{(\nabla\times\Field{v})}\,\dd{x},\\
b(\Field{w},\Field{v})&=\int_\Omega \mu \, \Field{w}\cdot 
\overline{\Field{v}}\,
\dd{x}.
\end{align}
}
With this we get  a weak formulation of 
Equations~(\ref{eigeneqn1}) and~(\ref{Bloch1}):
\begin{problem}\label{problem2}
Find $\omega^2\in\mathbb R$ and $\Field{H}\in \bhc$ such that 
\begin{equation}
\label{weak2}
a\left(\Field{H},\Field{v}\right)=\omega^2 \; b(\Field{H},\Field{v}) \quad
\forall\, \Field{v}\in \bhc,
\end{equation}
under the condition that
\begin{equation}
\label{ell_probl_for_proj}
b\left(\Field{H},\nabla p\right)=0 \qquad \forall p \in \bhp. 
\end{equation}
\end{problem}

The space $\bhc$ is the direct sum of the divergence-free subspace $\bhcP$
and the subspace  of gradient fields $\nabla p$, $p\in\bhp$.
Hence, $\Field{h}\in \bhc$ can be decomposed as 
$$\Field{h}=\Field{h}^\perp+\nabla p$$ (Helmholtz decomposition), where $p$ solves the 
equation
$$\int \nabla p\cdot \overline{\nabla v} \,\dd{x}= 
\int \Field{h} \cdot\overline{\nabla v}\,\dd{x} \quad \forall\,\,v\in\bhp.$$

\section{Finite Element Discretization}

It is crucial to inherit the properties of the Helmholtz decomposition
to the sub-spaces on the discrete level. 
Otherwise the discrete spectrum is polluted by many unphysical fields --
called spurious modes --
stemming from the space of gradient fields~\cite{Monk2003a}.
Using Nedelec's edge elements to discretize the space $\bhc$ 
and standard Lagrange elements of the same order to discretize the space $\bhp$
gives a discrete counterpart to the Helmholtz decomposition and to the divergence 
condition~\cite{Schmidt00a}.

We denote the discrete subspaces as follows:
$\bhch\subset\bhc$, $\bhph\subset\bhp$. 
Bloch periodicity is enforced
by a multiplication of basis functions associated with one of two 
corresponding periodic boundaries of the unit cell by the Bloch 
factor $\exp{(i \Kvec{k}\cdot\vec{a}_i)}$ (see Equation~\eqref{Bloch1}).
All interior basis functions remain unchanged. 
An alternative approach is discussed by Dobson et al\cite{Dobson2001a}:
In this approach a wave equation is formulated for the periodic part of 
the magnetic field, $\Field{u}(\vec{x})$ 
(see Equation~\eqref{Bloch1}). 
Modified finite element ansatz functions
are constructed from the Sobolev space $H\left(\curl+i\vec{k}\times\right)$.

The discretized problem corresponding to Problem \ref{problem2} reads as follows:
\begin{problem}
\label{problem3}
Find $\omega^2\in\mathbb R$ and $\Field{H}\in \bhch$ such that
\begin{equation}
\label{discreteeqn1}
a\left(\Field{H},\Field{\phi} \right)=\omega^2 \; b(\Field{H},\Field{\phi}) 
\quad\forall\Field{\phi}\in\bhch
\end{equation}
under the condition that
\begin{equation}
\label{discreteeqn2}
b\left(\Field{H},\nabla p\right)=0 \qquad \forall p \in \bhph. 
\end{equation}

\end{problem}

The finite element basis functions for 
$\bhch$ are denoted by $\Field{\phi_{j'}}$, ${1\leq j' \leq N_c}$
the basis functions for $\bhph$ by $\Field{\varphi_{k'}}$, ${1\leq k' \leq N_p}$. 
We expanding $\Field{H}$ in $\Field{\phi}_{i}$'s,
$\Field{H}=\sum u_i \Field{\phi}_{i}$.
Inserting into 
Equation~\eqref{discreteeqn1} 
yields the algebraic eigenvalue problem and the algebraic divergence condition
\begin{eqnarray}
\label{AlgebraicEigenvalueProblem}
\MyMatrix{A} \MyVector{u} &=& \lambda \MyMatrix{B} \MyVector{u}\\
\MyMatrix{G}^h\MyMatrix{B}\MyVector{u}&=&0
\end{eqnarray}
with $\MyMatrix{A}_{i,j}:= a\left(\Field{\phi}_{i}, \Field{\phi}_{j}\right)$,
$\MyMatrix{B}_{i,j}= b\left(\Field{\phi}_{i}, \Field{\phi}_{j}\right)$
and $\MyMatrix{G}_{i,j}$ defined by
$\nabla \Field{\varphi}_{i}=\sum_j\MyMatrix{G}_{j,i}\Field{\phi}_{j}$.
The matrix $\MyMatrix{A}$ is hermitian, positive semidefinite and $\MyMatrix{B}$ is hermitian, positive 
definite. 
In the algebraic form the Helmholtz decomposition reads as 
$\MyVector{u}=\MyVector{u}^\perp+\MyMatrix{G}\MyVector{p}$, where $\MyVector{p}$ solves the algebraic problem
\begin{equation}
\label{algebraichelmholtz}
\MyMatrix{G}^h\MyMatrix{B}\MyMatrix{G}\MyVector{p}=\MyMatrix{G}^h\MyMatrix{B}\MyVector{u}
\end{equation}

Due to the locality of the finite element basis functions all matrices are 
sparse.

\section{Numerical solution of the eigenvalue equation}
To solve the algebraic equation~\eqref{AlgebraicEigenvalueProblem} we use 
a preconditioned 
D\"ohler's method~\cite{Doehler1982a} which is based on minimizing the 
Rayleigh quotient.
To avoid that the iteration tumbles into the non-physical kernel of the 
$(\curl)$-operator we project the iterates onto the divergence-free subspace.

We use multi-level algorithms~\cite{Deuflhard2003a}  for preconditioning as well for performing 
the Helmholtz decomposition~\eqref{algebraichelmholtz}.
This is similar to the implementation by Hiptmair et al~\cite{Hiptmair2002a}.

With this, the computational time and the memory requirements
grow linearly with the number of unknowns~\cite{Burger2004a}.
Furthermore, we have implemented a residuum-based error 
estimator~\cite{Heuveline2001a}
and adaptive mesh refinement for 
the precise determination  of localized modes (see chapter~\ref{localizedchapter}).
As FE ansatz functions, we typically choose edge elements 
of quadratic order~\cite{Monk2003a}.

\section{Band structures of 3D photonic crystals}
\begin{figure}[htb]
\centering
\includegraphics[height=6cm]{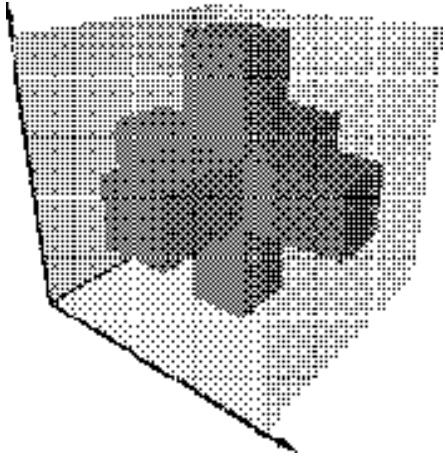}
\caption{Unit cell of a 3D photonic crystal ({\it scaffold}).
Bars with quadratic cross-sections intersect and form a 3D structure,
periodic boundary conditions apply to all pairs of opposing faces.}
\label{scaffold_fig1}
\end{figure}

A model problem for 3D photonic crystals are so-called 
{\it scaffold} structures~\cite{Dobson2000a}.
The geometry of a unit cell (sidelength $a$)
of a simple cubic lattice is shown in Fig.~\ref{scaffold_fig1}. 
It consists of bars (width $d=0.25\,a$) of a transparent material with
relative permittivity $\varepsilon_r=13$ and a background with $\varepsilon_r=1$ 
($\varepsilon=\varepsilon_r\varepsilon_0$, $\varepsilon_0$: free space permittivity).
For the calculation of the band structure the Bloch wavevector  $\Kvec{k}$
is varied along symmetry lines of the Brillouin zone ({\it cf.}~Dobson~\cite{Dobson2000a}).

\begin{figure}[hbt]
\centering
\psfrag{Band Structure - Scaffold structure}{}
\psfrag{(a)}{(a)}
\psfrag{(b)}{(b)}
\includegraphics[width=\textwidth]{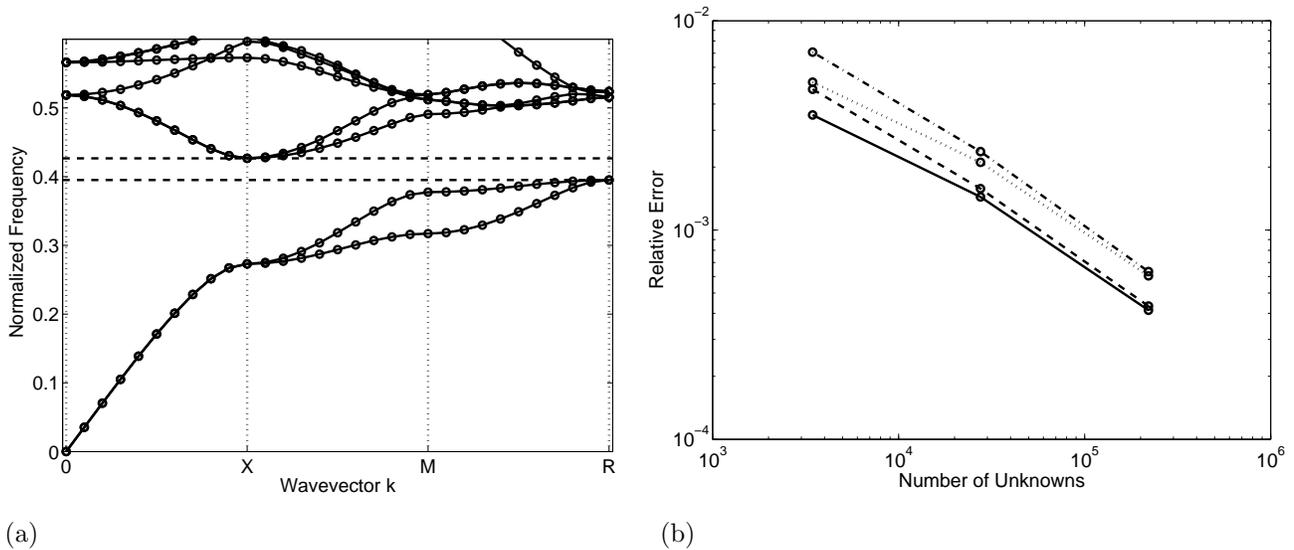}
\caption{(a) Band diagram for Bloch eigenmodes propagating in 
the scaffold structure. A complete bandgap is observed above $\tilde{\omega}\sim 0.4$.
(b) Convergence of the 
first four eigenvalues at the $X$-point towards the eigenvalues of the quasi-exact solutions.}
\label{scaffold_fig2}
\end{figure}

The band structure for light propagating in the scaffold structure
 is shown in Fig.~\ref{scaffold_fig2}a.
It exhibits a complete bandgap around the reduced frequency of
$\tilde{\omega}=\omega\,a/(2\pi\,c) \sim 0.4$ which is indicated by the dotted horizontal 
lines in Fig.~\ref{scaffold_fig2}a.
Table~\ref{scaffold_table} shows the four lowest
eigenvalues at the $X$-point($\Kvec{k}=(\pi/a,0,0)$) 
calculated on grids generated in 0, 1, resp.~2, uniform refinement steps from 
a coarse grid.
In each uniform refinement step, every tetrahedron is subdivided into eight new tetrahedra.
Shown are also the numbers of unknowns in the problem (number of ansatz functions in 
the finite element discretization) and typical computation times on a PC (Intel Pentium IV,
2.5\,GHz).
It can be seen that the computational effort grows linearly with the number of unknowns.
Figure~\ref{scaffold_fig2}\,(b) shows the dependence of the relative error of the 
four lowest eigenvalues ($|\omega_{i,N}-\omega_{i,q}|/\omega_{i,q}$) on the number 
of ansatz functions in the expansion of the eigenfunctions (number of unknowns).
Here, $\omega_{i,N}$ is the $i^{th}$ eigenfrequency of the discrete solution with 
N unknowns, $\omega_{i,q}$ is the  $i^{th}$ eigenfrequency of the quasi-exact 
solution obtained from a 
calculation on a finite-element grid with $N= 1764048$ unknowns.

\begin{table}
\begin{center}
\begin{tabular}{|rrrrrrr|}
\hline
Step & N$^o$ DOF & CPU time [min] & $\tilde{\omega}_1$& 
       $\tilde{\omega}_2$& $\tilde{\omega}_3$& $\tilde{\omega}_4$\\
\hline
 0 &     3450 &00:09.23  &2.736e-01 &2.740e-01 & 4.279e-01   & 4.288e-01 \\ 
 1 &    27572 & 01:46.33 &2.730e-01 &2.731e-01 & 4.266e-01   & 4.267e-01 \\ 
 2 &   220520 & 13:50.81 &2.728e-01 &2.728e-01 & 4.260e-01   & 4.260e-01 \\ \hline  
\end{tabular}
\end{center}
\caption{First eigenvalues of eigenmodes of the scaffold structure at $\Kvec{k}=X$.
Shown are the step number, the number of degrees of freedom of the problem, the CPU time (run on 
a standard PC), and the reduced frequencies of the four lowest eigenmodes.
}
\label{scaffold_table}
\end{table}

\begin{figure}[htb]
\centering
{\small (a)}
\includegraphics[height=5.5cm]{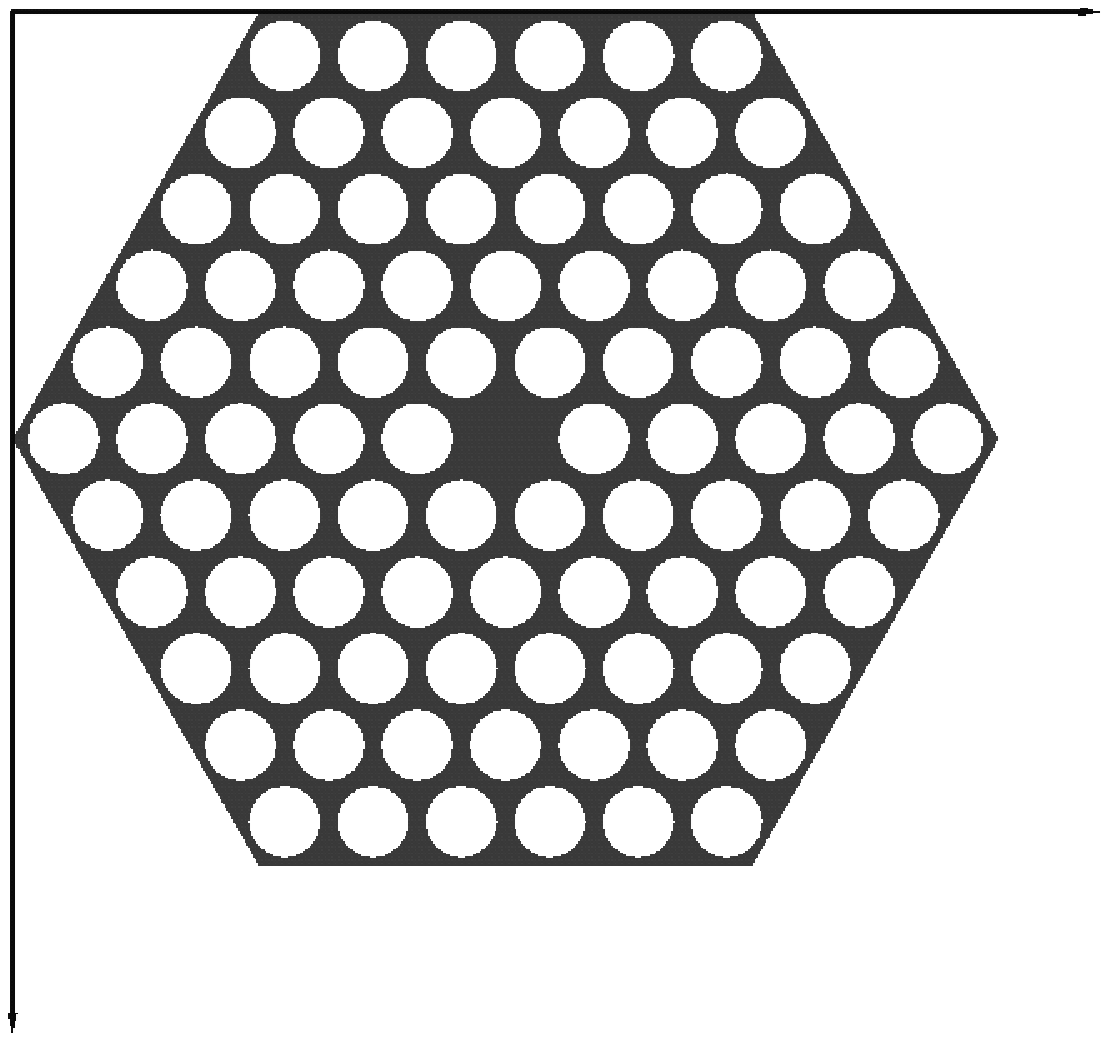}
{\small (b)}
\includegraphics[height=5.5cm]{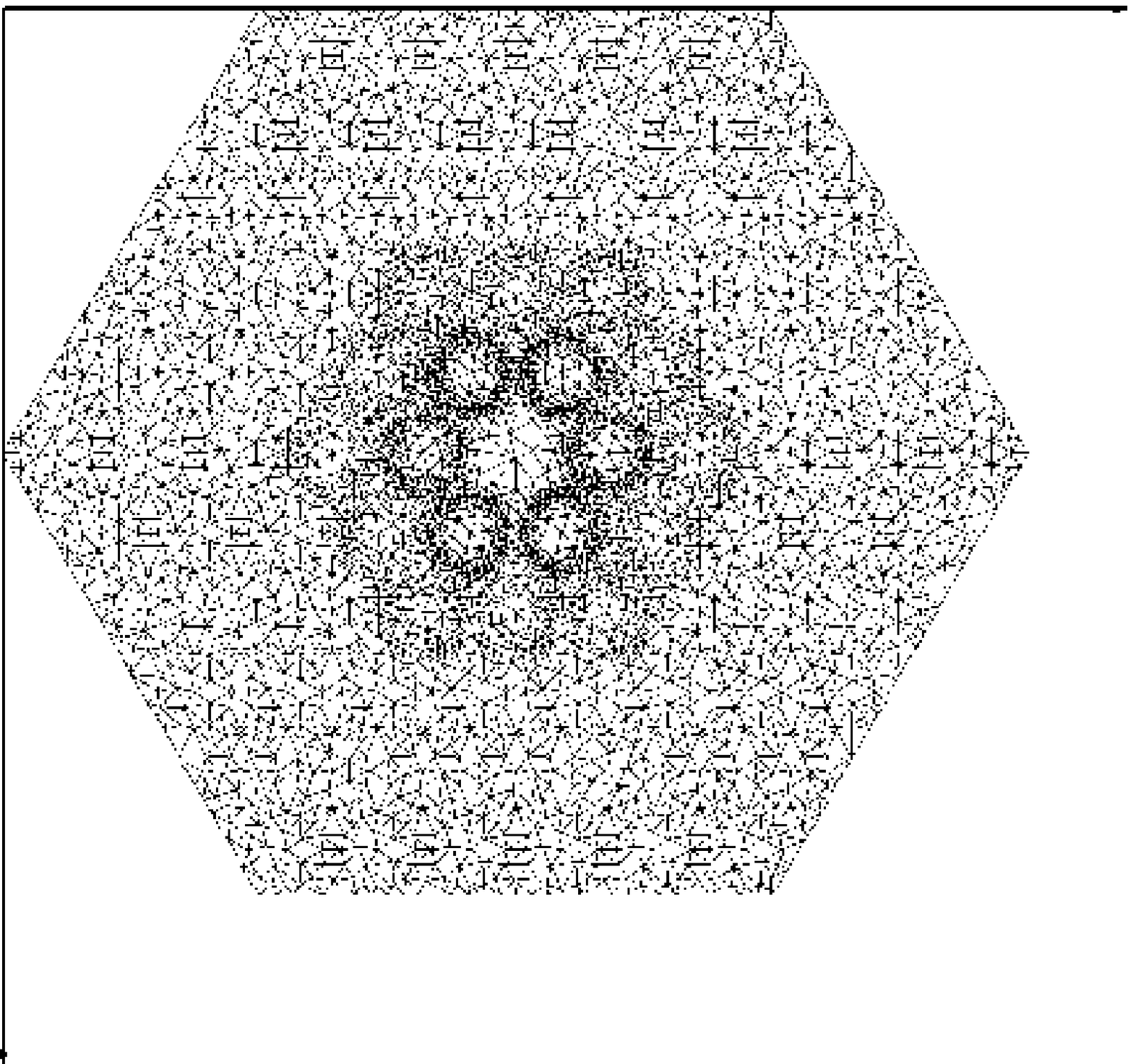}
\caption{Geometry (a) and coarse FE mesh (b) of a 2D photonic crystal structure 
with a central point defect.}
\label{pointdef_fig1}
\end{figure}

\begin{figure}[hbt]
\centering
{
{\small (a)}
\includegraphics[height=6.5cm]{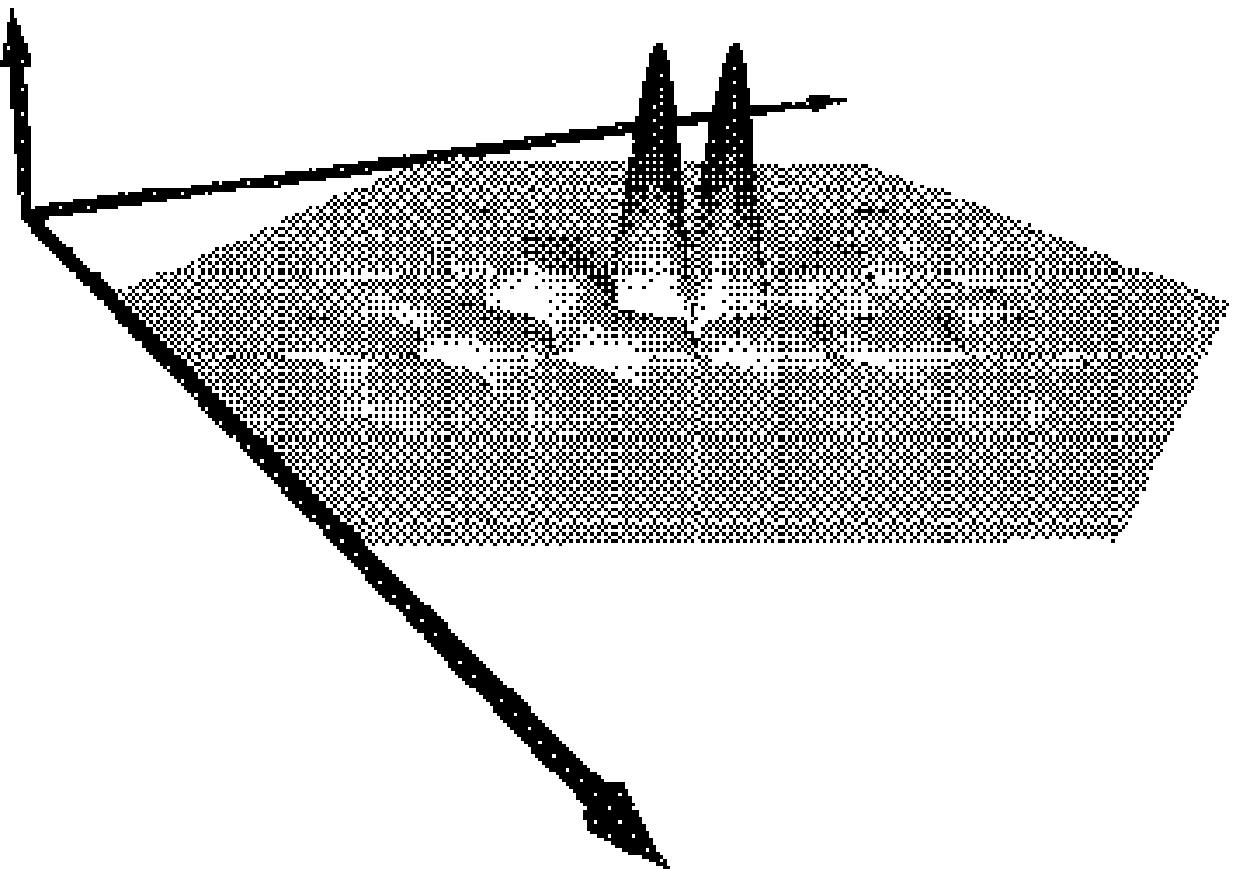}
{\small (b)}
\includegraphics[height=6.5cm]{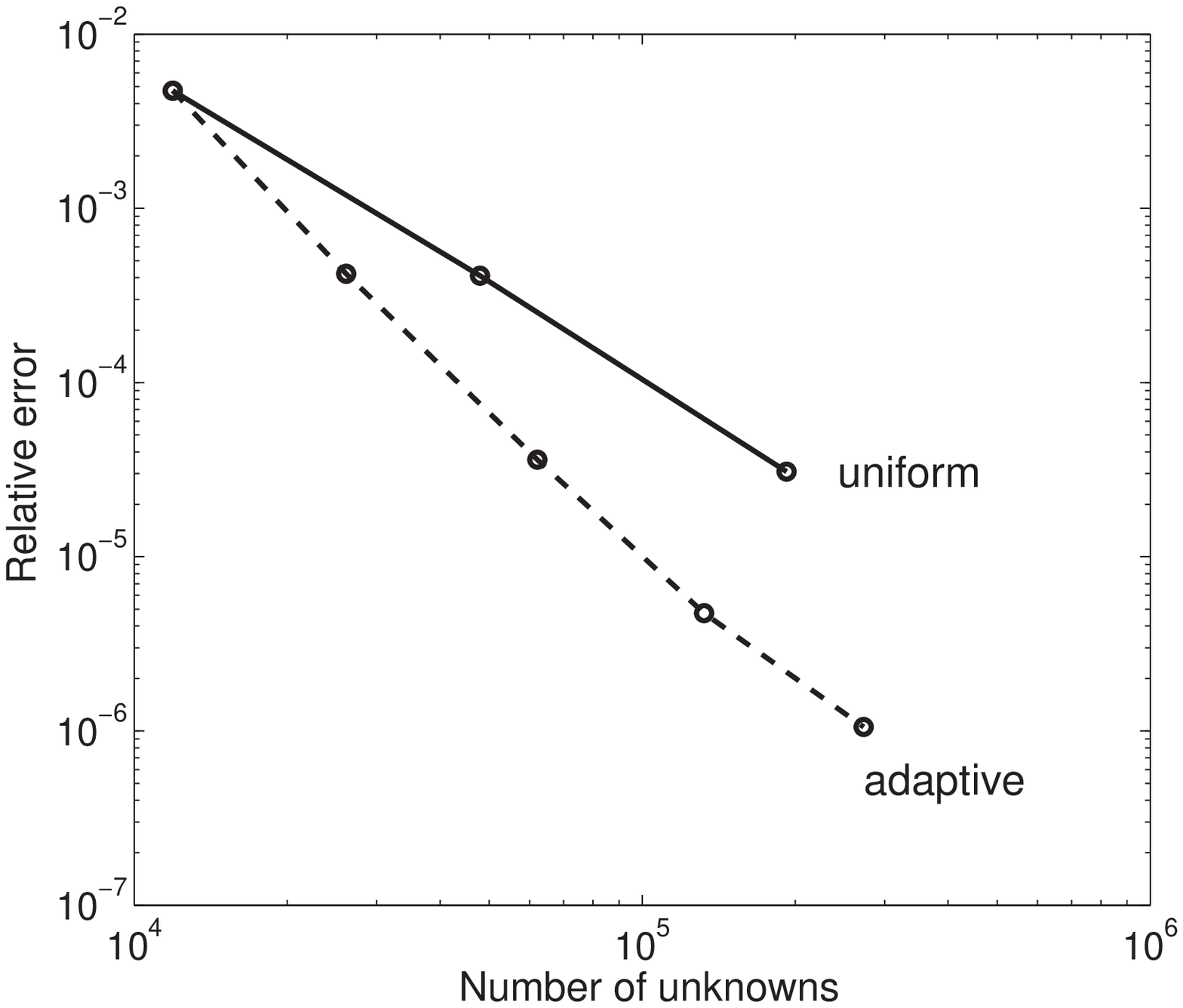}}
\caption{
(a) 
Distribution of the magnetic field intensity ($|\Field{H}(x,y)|$) for the 
lowest-frequency bound state at the point defect. 
(b) 
Comparison of the convergence of the eigenfrequency of the lowest frequency bound 
state towards a quasi-exact solution for adaptive and uniform refinement of the 
FE mesh.
}
\label{pointdef_fig2}
\end{figure}

\section{Calculation of defect modes using adaptive grid refinement}
\label{localizedchapter}
Light at a frequency inside the bandgap of a photonic crystal can be ``trapped'' inside
defects of the structure~\cite{Joannopoulos1995a}. 
This enables the construction of, e.g., waveguides (line defects) and micro-cavities
(point defects).

Figure~\ref{pointdef_fig1} (a) shows the geometry of a 2D photonic crystal with a point defect
(a missing pore in the center). It consists of a hexagonal lattice of 
air holes with a radius of $r=0.4\,a$ in 
a material with a relative electric permittivity of $\varepsilon_r=13$. 
A corresponding coarse triangular FE grid is shown in Figure~\ref{pointdef_fig1} (b). 
Please note that circular air pores are approximated by polygons. In the coarse grid 
shown in Figure~\ref{pointdef_fig1}\,(a) the pores close to the center are approximated 
to a higher accuracy than pores in the outer regions. 
Obviously, when refining the coarse grid in order to get a discrete solution which is 
closer to the solution of Problem~\ref{problem2} in some norm, the best strategy will 
not be to refine the grid uniformly, but to refine it in certain regions onlz. 
For this {\it adaptive} grid refinement we have implemented a residuum-based
error estimator~\cite{Heuveline2001a}.

Figure~\ref{pointdef_fig2} (a) shows the modulus of the magnetic field for the lowest-frequency
trapped eigenmode, computed with adaptive refinement of the FE mesh. 
Figure~\ref{pointdef_fig2} (b) 
shows the convergence of the eigenvalue
corresponding to this solution towards the eigenvalue of a quasi-exact solution
for adaptive grid refinement and for uniform grid refinement.
Obviously, adaptive grid refinement is especially useful when the 
sought solutions are geometrically localized, or
when the geometry exhibits sharp features, like discontinuities in the
refractive index distribution.
In this example, the use of the error estimator and adaptive refinement yields 
an order of magnitude in the accuracy of the error for a number of unknowns of 
$N\sim 10^{5}$.

\section{Photonic crystal slab waveguide}

Photonic crystal waveguides are promising candidates for a range of 
applications of micro- or nano-optical elements like dispersion 
compensators or input lines for further miniaturized elements. 
We examine a waveguiding structure composed  of a slab waveguide 
(confinement of the light in z-direction by a high-index guiding layer) 
combined with a 2D hexagonal array of air holes (PhC)~\cite{Michaelis2005a}. 
The waveguide is formed by a  defect  row of missing air holes in 
$\Gamma$-K-direction. 
Therefore, in certain wavelength ranges, the light is confined vertically 
by total internal reflection in the guiding layer and horizontally by the
photonic bandgap due to the 2D photonic crystal. 
We consider a guiding layer of height   
$z = 200nm$ and refractive index $n = 3.4$ with a substrate and superstrate of   
$z = 900nm$ each and refractive index $n = 1.45$, and six rows of pores
with refractive index $n=1.0$ on 
each side of the waveguide.
These parameters correspond to the material system of air pores in a 
$SiO_2 - Si - SiO_2$ slab structure.
The pore radius is $r \sim 0.36 a$, the lattice vectors have a length of $a=532\,$nm.

\begin{figure}[htb]
\centering
\psfrag{(a)}{\footnotesize mirror plane}
\psfrag{(b)}{\footnotesize air pore}
\psfrag{(c)}{\footnotesize guiding}
\psfrag{(d)}{\footnotesize layer}
\psfrag{(e)}{\footnotesize substrate}
\psfrag{(f)}{(a)}
\psfrag{(g)}{(b)}
\psfrag{x}{$x$}
\psfrag{y}{$y$}
\psfrag{z}{$z$}
\includegraphics[width=\textwidth]{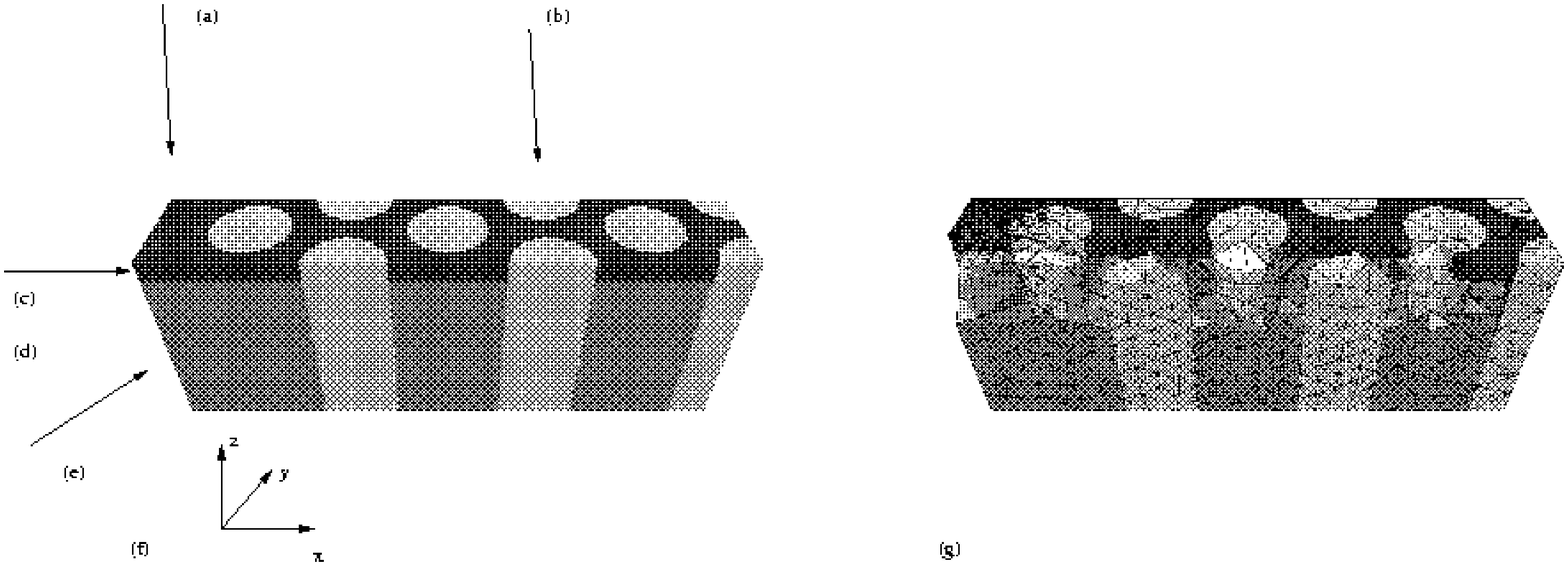}
\caption{(a) Geometry of the reduced unit cell of a W1 waveguide 
	     (dark gray: guiding layer, gray: substrate, light gray: cylindrical air pores). 
         (b) Visualization of the tetrahedral discretization of the geometry.}
\label{wg_geo_fig1}
\end{figure}

\begin{figure}[htb]
\centering
\psfrag{(a)}{(a)}
\psfrag{(b)}{(b)}
\includegraphics[height=9cm]{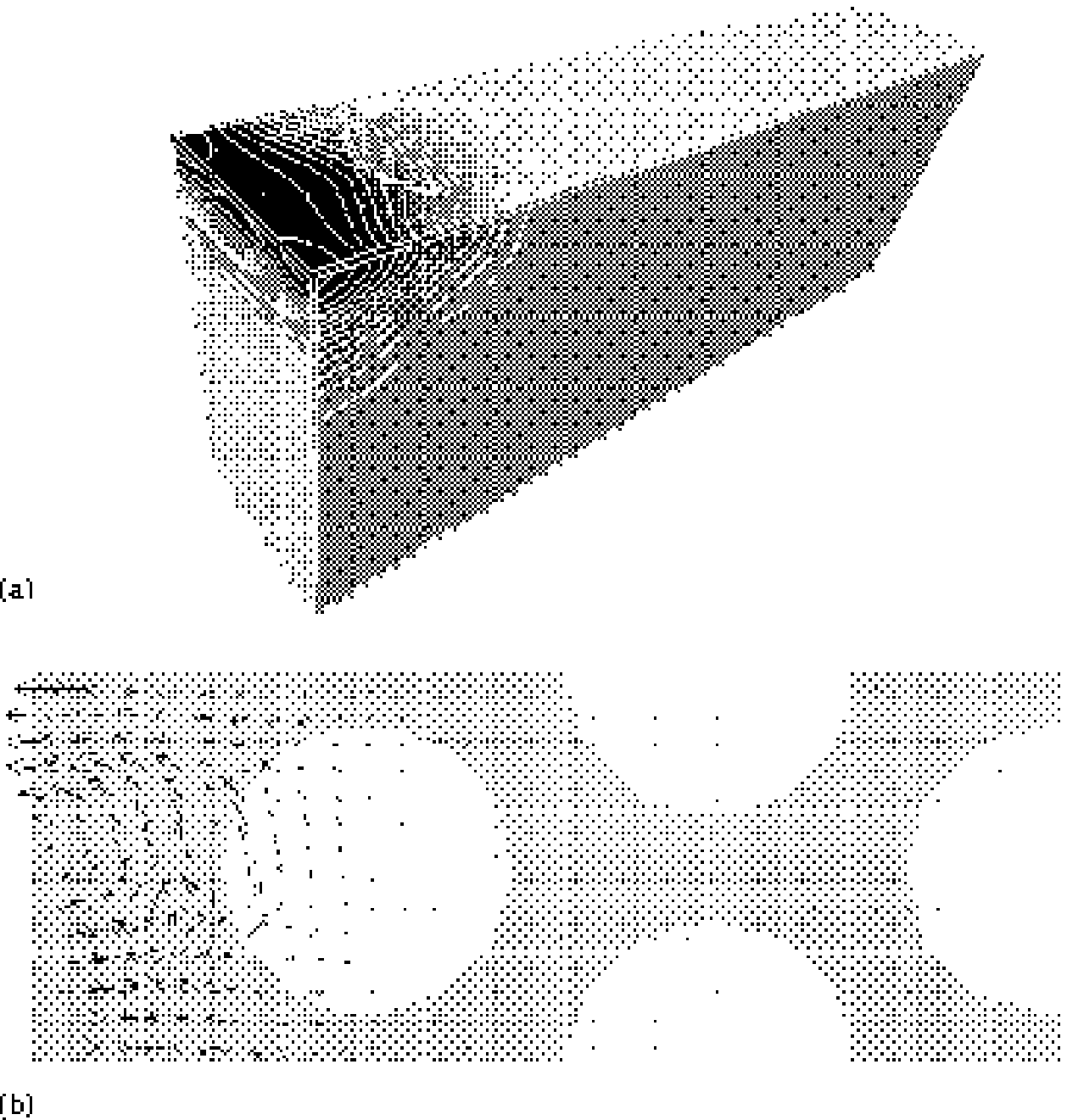}
\caption{(a) Magnitude of the electric field of a guided TE like mode in a gray 
		scale representation (black: high intensity).
		Iso-Intensity surfaces are indicated with white lines. 
         (b) Cross section of the $\vec{E}$ field distribution in the upper mirror plane 
             ($z=const.$). Additionally, the material distribution is indicated.}
\label{wg_sol_fig1}
\end{figure}
 
Due to the symmetry of the 
problem it is sufficient to restrict the computational domain to one quarter 
of a unit cell. This reduced computational domain is 
shown in Figure~\ref{wg_geo_fig1}(a). Here, the guiding layer is colored
in dark gray, the substrate in gray, and the air pores in light gray. 
Mirror symmetries are applied to the upper plane ($z$:
center of the guiding layer) and on the left ($x$ = 0). 
Periodic boundary conditions are applied to the front and back planes.
At the left of the computational domain a W1 waveguide (i.e., width $W=1.0\,a$)
is formed by a missing row of air pores. 

Fig.~\ref{wg_geo_fig1}(b) shows some of the tetrahedral elements in the discretization 
of the geometry. 
A typical coarse grid for this problem consists 
of about $10^4$ tetrahedra. 

Solving equation~\eqref{AlgebraicEigenvalueProblem} for this problem 
on a PC (Intel Pentium IV, 2.5 MHz, 2Gbyte RAM) typically takes our 
FEM code a time of about 2 min and delivers the eigenvalue and the complex vector field
for given Bloch wavevector $\vec{k}$. 
Figure~\ref{wg_sol_fig1} shows 
a specific guided mode in this structure.
Part (a) of this figure shows the amplitude of the magnetic field 
in a gray scale representation. 
The plotted solution has been calculated for a Bloch wavevector of 
$ka/(2 \pi) = 0.23$ and corresponds to an eigenvalue of  $\omega a/(2 \pi  c) \sim 0.326$, 
which lies inside the first bandgap of the (2D) photonic crystal. 
As can be seen from the figure, the light field is localized in the 
high index guiding layer and in the region of the missing pore. White lines
indicate equally spaced iso-intensity surfaces.
Part (b) of the figure shows a cross-section through the same (vectorial) solution in 
the upper mirror plane ($z=const.$). 
In this plane the electric field vectors are oriented in the $x-y$-plane, 
this solution corresponds to a TE-like mode. 

\section{Transmission through a finite photonic crystal}

In order to simulate the transmission of an incident light field $\Field{u}_{in}$ with a frequency $\omega$
through a photonic crystal of finite width we perform scattering calculations.
In this case we have to seek a solution $\Field{u}$ which fulfills Equation~\eqref{eigeneqn1} 
resp.~\eqref{eigeneqn1e} for the given frequency $\omega$ on the computational 
domain~$\Omega$ with the following boundary condition:
(a) the field on the boundary can be written as a superposition of incident and 
scattered light field: $\Field{u}=\Field{u}_{in}+\Field{u}_{sc}$,
(b) the scattered field is purely outgoing ({\it radiation condition}), 
which gives a condition for the outward normal derivative
of $\Field{u}$, $\partial_\nu\Field{u}_{sc}$. 
We obtain a relation between $\partial_\nu\Field{u}_{sc}$ and $\Field{u}_{sc}$ by 
using a modified type of perfectly matched layer boundary conditions, which 
also allows to treat certain types of {\it inhomogeneous} exterior 
domains~\cite{Schmidt02H}.

\begin{figure}[htb]
\centering
\psfrag{(a)}{ plane}
\psfrag{(b)}{ wave}
\psfrag{(c)}{ Si}
\psfrag{(d)}{ air pore}
\psfrag{(alpha)}{$\alpha$}
\psfrag{(bc1)}{\footnotesize (BC t1)}
\psfrag{(bc2)}{\footnotesize (BC p1)}
\psfrag{(bc3)}{\footnotesize (BC t2)}
\psfrag{(bc4)}{\footnotesize (BC p2)}
\psfrag{x}{x}
\psfrag{y}{y}
\includegraphics[width=0.7\textwidth]{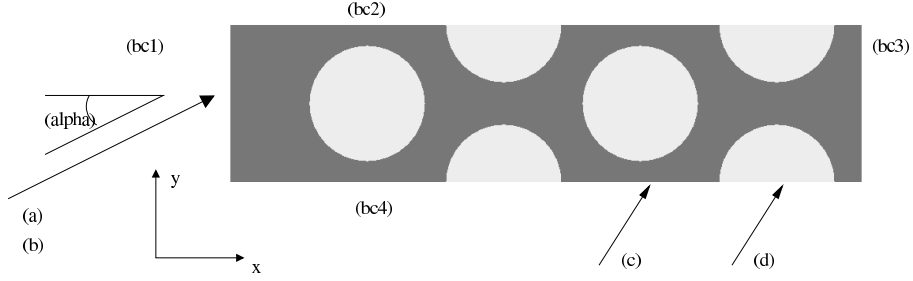}
\caption{Geometry of a 2D finite photonic crystal: Bloch-periodic boundary conditions apply to the 
boundaries (BC p1) and (BC p2), transparent boundary conditions apply to the boundaries 
(BC t1) and (BC t2),
a plane wave is incident onto the boundary (BC t1).}
\label{trans_geo}
\end{figure}

\begin{figure}[h!]
\centering
\psfrag{(a)}{(a)}
\psfrag{(b)}{(b)}
\psfrag{(c)}{(c)}
\psfrag{(d)}{(d)}
\psfrag{(e)}{(e)}
\psfrag{(f)}{(f)}
\psfrag{(g)}{(g)}
\psfrag{(h)}{(h)}
\psfrag{(i)}{(i)}
\psfrag{(j)}{(j)}
\psfrag{x}{x}
\psfrag{y}{y}
\psfrag{z}{z}
\psfrag{k(in)}{$\vec{k}_{in}$}
\includegraphics[width=\textwidth]{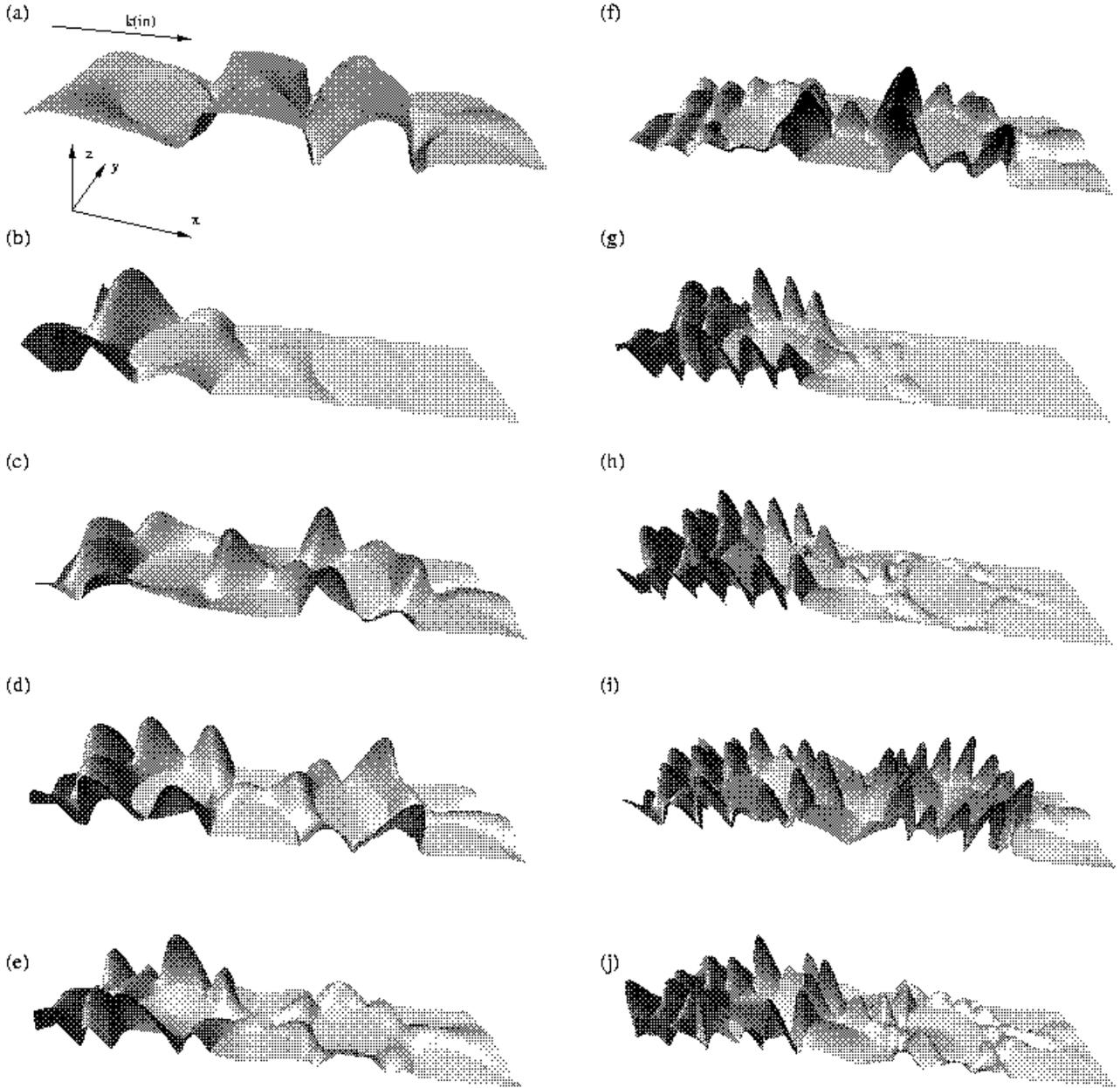}
\caption{
Light fields propagating through a finite 2D photonic crystal. 
Plotted is the magnitude of the magnetic field, $|\Field{H}_z(x,y)|$,
in a gray scale representation.
The direction of the plane wave is indicated by the vector 
$\vec{k}_{in}$ which is incident under an angle of $\alpha=10\,deg$
(compare Fig.~\ref{trans_geo}).
The different plots correspond to incident plane waves with different 
frequencies: 
(a): $\tilde{\omega}=\omega a/2\pi c = 0.182$,
(b): $\tilde{\omega}=0.303$, 
(c): $\tilde{\omega}=0.385$, 
(d): $\tilde{\omega}=0.4$, 
(e): $\tilde{\omega}=0.5$, 
(f): $\tilde{\omega}=0.625$, 
(g): $\tilde{\omega}=0.667$, 
(h): $\tilde{\omega}=0.714$, 
(i): $\tilde{\omega}=0.741$, 
(j): $\tilde{\omega}=0.769$. 
}
\label{trans_sol}
\end{figure}

\begin{figure}[htb]
\centering
\psfrag{(a)}{(a)}
\psfrag{(b)}{(b)}
\includegraphics[width=\textwidth]{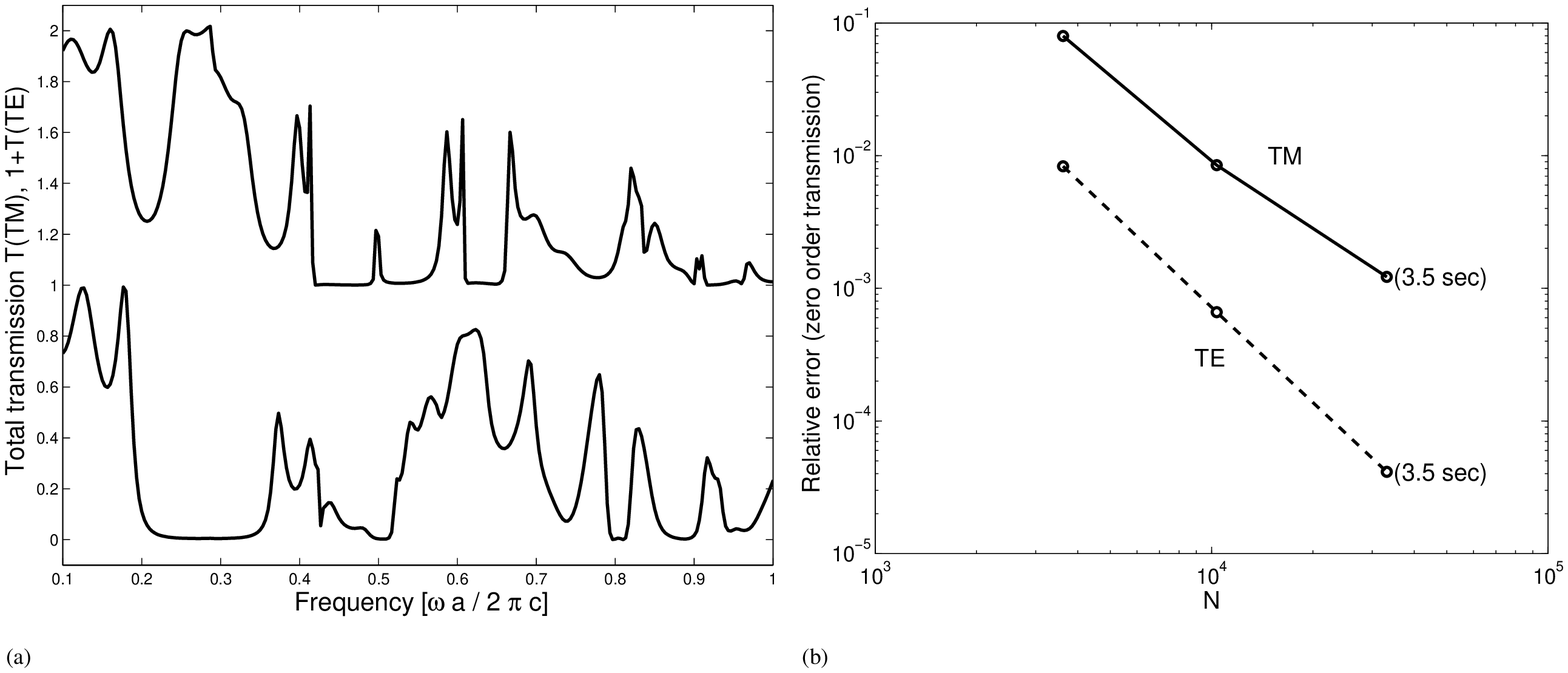}
\caption{(a): Transmission of TE and TM light fields through a finite 2D photonic 
crystal for an angle of incidence of $\alpha=0\,deg$.
(b): Convergence of the relative error of the zero order transmission for TE and 
TM light fields (for a frequency of $\tilde{\omega}\sim0.37$ and uniform grid refinement).
Cpu times for computations on a standard laptop are indicated (Intel Pentium IV, 2.0~GHz).
}
\label{trans_conv_fig}
\end{figure}

In the following example we examine the 
light transmission through a finite 2D photonic crystal consisting of four rows 
of air pores in a high index material.
Figure~\ref{trans_geo} shows the geometry of the problem:
Air pores ($n=1.0$) of radius $r=0.366\,a$ in a high-index material ($n=3.4$) 
form a finite hexagonal pattern.
Periodic boundary conditions apply to two boundaries  of the computational 
domain.
Transparent boundary conditions to the exterior (which is assumed to be also 
filled by the high index material) are applied to the two other facets of the 
computational window.
A plane wave with a freely chosen angle of incidence $\alpha$ is incident onto  
one boundary.
In order to measure the field transmission through the domain, the 
field on the opposing facet (BC t2) detected.

Figure~\ref{trans_sol} shows calculated field distributions for different 
frequencies $\tilde{\omega}=\omega a/2\pi c=a/\lambda $ of the incident plane wave
($\alpha=10$\,deg).
Each of these solutions has been calculated using adaptive mesh refinement,
finite elements of quadratic 
order, and typically $N\sim 4\cdot 10^4$ unknowns.
The total calculation time  on a standard laptop (Intel Pentium IV, 2.0~GHz) for each
solution amounts about 10\,sec.
The fields in Figure~\ref{trans_sol} (a)-(j) correspond to decreasing wavelength.
It can easily seen how dramatically the transmission is changes with the wavelength:
Certain wavelengths (b, g, h, j) correspond to (partial or full) band-gaps of 
the photonic crystal, where light transmission is suppressed. At other 
wavelengths the light is transmitted; resonance behavior / slow group 
velocities can be discovered by the observed increased field amplitudes in 
the regions between the air holes, see e.g.~(f). 

We decompose the field at the facets of the computational window into 
Fourier series:
\newline
$u(y)=\sum_{n=-\infty}^\infty A_n \exp(i 2\pi n y / L)$.
The relative transmission through the photonic crystal
is then given by 
\begin{equation}
T=\frac{\sum_{|\vec{k}_n|<|\vec{k}_{in}|}\sin(\vec{k}_n, \vec{n})A_n^2}{A_{in}^2}\quad,
\end{equation}
where $\vec{n}$ is the normal vector on the end facet.
Figure~\ref{trans_conv_fig}\,(a) shows the transmission in dependence on the 
frequency of the incident light for an angle of incidence $\alpha=0\,deg$. 
Detailed structures corresponding to bands and band gaps can be observed.
Figure~\ref{trans_conv_fig}\,(b) shows the relative error of the lowest order 
Fourier coefficient of the transmitted light field in dependence on the number 
of ansatz functions. 
For TE light fields ($|\Field{E}_z(x,y)|$), the errors are lower due to the smoothness of the electric 
field in this case. 
However, even for TM light fields ($|\Field{H}_z(x,y)|$) very accurate transmission 
coefficients with errors in the $10^{-3}$-range can be gained with rather low 
numbers of unknowns and in short computation times on standard PC's.

\newpage
\section{Conclusion}
In this paper we have presented an adaptive finite-element method solver for the computation of 
electromagnetic eigenmodes and scattered light fields. 
The convergence analysis of solutions for model problems
shows the efficiency of the methods.
Our solver has been shown to give very accurate solutions for typical 
problems arising in nano- and micro-optics -- even challenging 3D design tasks 
can be tackled on standard personal computers.

\acknowledgements
We thank P.~Deuflhard, R.~M\"arz, D.~Michaelis, and C.~W\"achter for fruitful discussions,
and we acknowledge support by  the initiative DFG Research Center {\sc Matheon} 
of the Deutsche Forschungsgemeinschaft, DFG,  and by the German Federal Ministry of
Education and Research, BMBF, under contract No.~13N8252 ({\sc HiPhoCs}).

\bibliography{/home/numerik/bzfburge/texte/biblios/phcbibli,/home/numerik/bzfburge/texte/biblios/group04}   
\bibliographystyle{spiebib}   

\end{document}